\documentclass{webofc}
\usepackage[varg]{txfonts}   
\begin{document}
\title{ Wigner function and the probability representation of quantum
states}
%
%

\author{Margarita A. Man'ko\inst{1}\fnsep\thanks{\email{mmanko@sci.lebedev.ru}} \and
        Vladimir I. Man'ko\inst{1,2}\fnsep\thanks{\email{manko@sci.lebedev.ru}}}

\institute{ P. N. Lebedev Physical Institute, Russian Academy of
Sciences, Leninskii Prospect 53, Moscow 119991, Russia \and Moscow
Institute of Physics and Technology (State University),
Institutski\'{\i} per. 9, Dolgoprudny\'{\i}, Moscow Region 141700,
Russia }

\abstract{%
The relation of the Wigner function with the fair probability
distribution called tomographic distribution or quantum tomogram
associated with the quantum state is reviewed. The connection of the
tomographic picture of quantum mechanics with the integral Radon
transform of the Wigner quasidistribution is discussed. The
Wigner--Moyal equation for the Wigner function is presented in the
form of kinetic equation for the tomographic probability
distribution both in quantum mechanics and in the classical limit of
the Liouville equation. The calculation of moments of physical
observables in terms of integrals with the state tomographic
probability distributions is constructed having a standard form of
averaging in the probability theory. New uncertainty relations for
the position and momentum are written in terms of optical tomograms
suitable for direct experimental check. Some recent experiments on
checking the uncertainty relations including the entropic
uncertainty relations are discussed. }
\maketitle
\section{Introduction}
\label{intro}
The states of quantum systems are identified with the wave
function~\cite{Schr26} or the density
matrix~\cite{Landau,vonNeumann}. For the quantum particle, in 1932
the function $W(q,p)$ was introduced by Wigner~\cite{Wigner32}; this
function contains all information on the state and is similar to the
classical probability density $f(q,p)$ in the phase space. The
Wigner function can take negative values, so it is not a a fair
probability distribution. Nevertheless, using the invertible Radon
transform~\cite{Radon17}, one can obtain the fair probability
distribution~\cite{BerBer} called the optical tomogram measured in
quantum-optics experiments~\cite{Raymer}. In \cite{Mancini96}, it
was suggested to identify the quantum states with
tomographic-probability distributions as primary objects which are
alternatives to the wave functions or the density matrices.

The aim of this work is to present a review of the approach (see
also \cite{RitaFP,IbortPS,NuovoCim}) and obtain new quantum
inequalities associated with the tomographic probabilities and
Wigner function.

\section{Tomographic probability distributions}
\label{sec-1}
The Wigner function is determined by the density operator $\hat\rho$
\begin{equation}\label{1.6}
W(q,p)=2\,\mbox{Tr}\,\big(\hat\rho\hat{\cal D}(2\alpha)\hat
I\big),\qquad \alpha=(q+ip)/\sqrt2,
\end{equation}
where $\hat{\cal D}(2\alpha)=\exp\big(2\alpha\hat
a^\dagger-2\alpha^*\hat a\big)$, $\big[\hat a,\hat
a^\dagger\big]=1$, and the parity operator $\hat I$ is $\hat
I\psi(x)=\psi(-x)$.

The inverse transform reads
\begin{equation}\label{1.7}
\hat\rho=\pi^{-1}\int W(q,p)\hat{\cal D}(2\alpha)\hat I\,dq\,dp.
\end{equation}
The optical tomogram is given by the Radon transform~\cite{Radon17}
of the Wigner function
\begin{equation}\label{1.8}
w(X,\theta)=\int
\delta(X-q\cos\theta-p\sin\theta)W(q,p)\,\frac{dq\,dp}{2\pi}\,.
\end{equation}
The symplectic tomogram $w(X,\mu,\nu)$ in terms of the Wigner
function and optical tomogram reads
\begin{equation}\label{1.9}
w(X,\mu,\nu)=\int \delta(X-\mu q-\nu
p)W(q,p)\,\frac{dq\,dp}{2\pi}=\frac{1}{\sqrt{\mu^2+\nu^2}}\,
w\left(\frac{X}{\sqrt{\mu^2+\nu^2}},
\mbox{tan}^{-1}\frac{\nu}{\mu}\right).
\end{equation}
The density operator in terms of the symplectic tomogram is
\begin{equation}\label{1.10}
\hat\rho=(2\pi)^{-1}\int w(X,\mu,\nu)\exp\big[i(X-\mu\hat q-\nu\hat
p)\big]\,dX\,d\mu\,d\nu.
\end{equation}
Recall that tomograms are normalized probability distributions,
i.e., $w(X,\theta)\geq 0$,  $w(X,\mu,\nu)\geq 0$, $\int
w(X,\theta)\,dX=1$, and $\int w(X,\mu,\nu)\,dX=1$.

The von Neumann equation for the density operator
\begin{equation}\label{1.11}
\frac{\partial\hat\rho}{\partial t}+i\big[\hat
H,\hat\rho\big]=0,\qquad \hat H=\frac{\hat p^2}{2}+\hat U
\end{equation}
was written for the optical tomogram in \cite{KorenJRLR,AmKorenPRA}
as follows:
\begin{eqnarray}
\frac{\partial}{\partial t}w(X,\theta,t)=
\left[\cos^2\theta\,\frac{\partial}{\partial\theta}
-\frac{1}{2}\sin2\theta\left\{1+X\frac{\partial}{\partial X}\right\}
\right]w(X,\theta,t) \nonumber \\
+2\left[\mbox{Im}~U\left\{
\sin\theta\,\frac{\partial}{\partial\theta}
\left[\frac{\partial}{\partial X}\right]^{-1}
+X\cos\theta+i\,\frac{\sin\theta}{2} \frac{\partial}{\partial
X}\right\}\right] w(X,\theta,t).
        \label{eq35}
\end{eqnarray}
In the classical limit, this equation converts into the Liouville
equation for classical optical tomogram $w_{\rm cl}(X,\theta,t)=\int
\delta(X-q\cos\theta-p\sin\theta)f(q,p,t)\,dq\,dp$, where $f(q,p,t)$
is the probability density in the phase space,
\begin{eqnarray}
\frac{\partial}{\partial t}w_{\rm cl}(X,\theta,t)=
\left[\cos^2\theta\frac{\partial}{\partial\theta} -\frac{1}{2}\sin
2\theta\left\{1+X\frac{\partial}{\partial X}\right\}
\right]w_{\rm cl}(X,\theta,t) \nonumber \\
+\left[\frac{\partial}{\partial q}\,U\left\{
q\rightarrow\sin\theta\frac{\partial}{\partial\theta}
\left[\frac{\partial}{\partial X}\right]^{-1} +X\cos\theta\right\}
\sin\theta\frac{\partial}{\partial X}\right] w_{\rm cl}(X,\theta,t),
        \label{eq54}
\end{eqnarray}
and for classical symplectic tomogram $M_{\rm cl}(X,\mu,\nu,t)$ it
reads
\begin{eqnarray}
\frac{\partial}{\partial t}M_{\rm cl}(X,\mu,\nu,t)=
\mu\frac{\partial}{\partial\nu}
M_{\rm cl}(X,\mu,\nu,t) 
+\left[\frac{\partial}{\partial q}U
\left\{q\rightarrow-\left[\frac{\partial}{\partial X}\right]^{-1}
\frac{\partial}{\partial\mu}\right\} \nu\frac{\partial}{\partial
X}\right] M_{\rm cl}( X,\mu,\nu,t).
\label{eq55}
\end{eqnarray}

The statistical properties of the position and momentum are
expressed in terms of the optical tomogram as follows:
\begin{equation}\label{1.14}
\langle\hat q^n\rangle=\mbox{Tr}\,\hat\rho\hat q^n=\int
w(X,\theta=0)X^ndX, \qquad \langle\hat
p^n\rangle=\mbox{Tr}\,\hat\rho\hat q^n=\int w(X,\theta=\pi/2)X^ndX.
\end{equation}

\section{Quantum uncertainty relations in terms of tomograms and Wigner function}
\label{sec-2}
The Heisenberg uncertainty relation (at $\hbar=1$) in the form
\begin{eqnarray}\label{1.15}
&&\left[\int w(X,\theta=0)X^2dX-\left(\int
w(X,\theta=0)X\,dX\right)^2\right]\nonumber\\
&&\times \left[\int w(X,\theta=\pi/2)X^2dX-\left(\int
w(X,\theta=\pi/2)X\,dX\right)^2\right]\geq\frac{1}{4}
\end{eqnarray}
has been checked in \cite{PorzioPS,BelliniPRA}.

The optical tomogram satisfies the entropic
inequality~\cite{Fedele,RitaFP}
\begin{equation}\label{1.1}
\ln\pi e+\int w(X,\theta)\ln w(X,\theta)\,dX+ \int
w(X,\theta+\pi/2)\ln w(X,\theta+\pi/2)\,dX\leq 0.\end{equation} This
inequality was checked in \cite{BelliniPRA}.

To derive another inequality, we introduce four numbers $p_1$,
$p_2$, $p_3$, and  $p_4$
\begin{equation}\label{1.2}
p_1=\int_{-\infty}^{x_1}w(X,\theta)\,dX,\quad
p_2=\int_{x_1}^{x_2}w(X,\theta)\,dX,\quad
p_3=\int_{x_2}^{x_3}w(X,\theta)\,dX,\quad
p_4=\int_{x_3}^{\infty}w(X,\theta)\,dX,
\end{equation}
where $-\infty<x_1\leq x_2\leq x_3<\infty.$ Then one has the
inequality which is an analog of the subadditivity condition for
bipartite system
\begin{eqnarray}\label{1.3}
&&-p_1\ln p_1-p_2\ln p_2-p_3\ln p_3-p_4\ln p_4\leq-(p_1+p_2)\ln(p_1+p_2)\nonumber\\
&&-(p_3+p_4)\ln(p_3+p_4)
-(p_1+p_3)\ln(p_1+p_3)-(p_2+p_4)\ln(p_2+p_4).
\end{eqnarray}

The new inequality for the Wigner function $W(q,p)$ of the pure
state can be also found. If one has four numbers, which are
functionals of the Wigner function of the form
\begin{eqnarray}\label{1.4}
\Pi_1=\int_{-\infty}^{x_1}\hspace{-1.2mm}\int_{-\infty}^{\infty}
W^2(q,p)\,\frac{dq\,dp}{2\pi}\,,\qquad
\Pi_2=\int_{x_1}^{x_2}\hspace{-2mm}\int_{-\infty}^{\infty}
W^2(q,p)\,\frac{dq\,dp}{2\pi}\,,\nonumber\\
\\
\Pi_3=\int_{x_2}^{x_3}\hspace{-2mm}\int_{-\infty}^{\infty}
W^2(q,p)\,\frac{dq\,dp}{2\pi}\,,\qquad
\Pi_4=\int_{x_3}^{\infty}\hspace{-2mm}\int_{-\infty}^{\infty}
W^2(q,p)\,\frac{dq\,dp}{2\pi}\,,\nonumber
\end{eqnarray}
an inequality analogous to (\ref{1.3}) is valid, namely,
\begin{eqnarray}\label{1.5}
&&-\Pi_1\ln\Pi_1-\Pi_2\ln\Pi_2-\Pi_3\ln\Pi_3-\Pi_4\ln\Pi_4\leq-(\Pi_1+\Pi_2)\ln(\Pi_1+\Pi_2)\nonumber\\
&&-(\Pi_3+\Pi_4)\ln(\Pi_3+\Pi_4)
-(\Pi_1+\Pi_3)\ln(\Pi_1+\Pi_3)-(\Pi_2+\Pi_4)\ln(\Pi_2+\Pi_4).
\end{eqnarray}

Inequality~(\ref{1.3}) can be checked experimentally. Optical
tomograms $w(X,\theta)$ of photon states are measured by homodyne
detector~\cite{BelliniPRA}. They must satisfy inequality~(\ref{1.3})
for an arbitrary local oscillator phase $\theta$ and arbitrary
numbers $x_1$, $x_2$, and $x_3$.

\section{Conclusions}
\label{sec-3}
To conclude, we point out our main new results.

We obtained new inequalities for optical tomograms~(\ref{1.2}) and
(\ref{1.3}), which can be measured experimentally. Also we found new
integral inequalities for the Wigner function $W(q,p)$, which can be
checked in the experiments similar to the ones performed in
\cite{Raymer}, where the Wigner function of photon states is
reconstructed from homodyne detection. The inequalities we obtained
are analogous to the subadditivity condition for entropy of
bipartite systems, but they are valid for systems without
subsystems. Such kinds of inequalities were recently discussed in
\cite{RitaPST-CEWQO,OlgaJRLR14}. The entropic
inequalities~\cite{RitaPS2012} correspond to the general properties
of nonnegative-number sets~\cite{RitaJRLR2013}.

\section*{Acknowledgements}
The authors are grateful to the Organizers of the Wigner~111
Scientific Symposium (11--13 November 2013, Budapest, Hungary) for
invitation and kind hospitality.


\begin{thebibliography}{}

\bibitem{Schr26}
E. Schr\"{o}dinger, Ann. Phys. \textbf{79}, 361 (1926)


\bibitem{Landau} L. D. Landau, 
 Z. Phys. {\bf 45}, 430  
(1927)

\bibitem{vonNeumann}
J. von Neumann, Nach. Ges. Wiss. G\"{o}ttingen, \textbf{11}, 245
(1927) 

\bibitem{Wigner32}
E. Wigner,  Phys. Rev. {\bf 40}, 749 (1932)

\bibitem{Radon17}
J. Radon, Ber. Verh. Sachs. Akad. \textbf{69}, 262 (1917)

\bibitem{BerBer}
J. Bertrand and P. Bertrand, Found. Phys. {\bf 17}, 397 (1987)


\bibitem{Raymer}
D. T. Smithey, M. Beck, M. G. Raymer, and A. Faridani, Phys. Rev.
Lett. {\bf 70}, 1244 (1993)

\bibitem{Mancini96}
S. Mancini, V. I. Man'ko, and P. Tombesi,  Phys.~Lett.~A
\textbf{213}, 1 (1996)

\bibitem{RitaFP}
M. A. Man'ko and V. I. Man'ko, Found.~Phys.
{\bf 41}, 330 
(2011) 

\bibitem{IbortPS}
A. Ibort, V. I. Man'ko, G. Marmo, A. Simoni, and F. Ventriglia,
Phys. Scr. {\bf 79}, 065013 (2009)

\bibitem{NuovoCim}
M. A. Man'ko, V. I. Man'ko, G. Marmo, A. Simoni, and F. Ventriglia,
Nuovo Cim. C {\bf 36}, Ser.~3, 163 (2013)

\bibitem{KorenJRLR}
Ya. A. Korennoy and V. I. Man'ko, J. Russ. Laser Res. \textbf{32},
74 (2011)

\bibitem{AmKorenPRA}
G. G. Amosov, Ya. A. Korennoy, and V. I. Man'ko,  
Phys. Rev. A {\bf 85}, 052119 (2012)

\bibitem{PorzioPS}
V. I. Man'ko, G. Marmo, A. Porzio, S. Solimeno, and F. Ventriglia,
Phys. Scr. {\bf 83}, 045001 (2013)

\bibitem{BelliniPRA}
M. Bellini, A. S. Coelho, S. N. Filippov, V. I. Man'ko, and A.
Zavatta, 
Phys. Rev. A {\bf 85}, 052129 (2012)

\bibitem{Fedele}
S. De Nicola, R. Fedele, M. A. Man'ko, and V.I. Man'ko,
Acta Phys. Hung. Sect. B: Quantum Electron. {\bf 20}, 
261  
(2004)

\bibitem{RitaPST-CEWQO}
M. A. Man'ko and V. I. Man'ko, ``Quantum strong subadditivity
condition for systems without subsystems,''  arXiv:1312.6988
[quant-ph]; Phys. Scr. T (2014, in press)

\bibitem{OlgaJRLR14}
V. N. Chernega and O. V. Man'ko, J. Russ. Laser Res. \textbf{35}, 27
(2014)

\bibitem{RitaPS2012}
M. A. Man'ko and V. I. Man'ko, Phys. Scr. T \textbf{147}, 014020
(2012)

\bibitem{RitaJRLR2013}
M. A. Man'ko and V. I. Man'ko, J. Russ. Laser Res. \textbf{34}, 203
(2013)


\end{thebibliography}

\end{document}